\documentclass[runningheads]{svmult}

\usepackage{makeidx}   % allows index generation
\usepackage{graphicx}  % standard LaTeX graphics tool
\usepackage{subeqnar}  % subnumbers individual equations
\usepackage{multicol}  % used for the two-column index
\usepackage{physprbb}  % modified textarea for proceedings,
\makeindex             % used for the subject index
\begin{document}
\title*{The Evolution of Circumstellar Disks:\protect\ Lessons from the VLT and ISO}
\toctitle{The Evolution of Circumstellar Disks:\protect\newline Lessons from the VLT and ISO}
% allows explicit linebreak for the table of content
%
%
\titlerunning{Evolution of Circumstellar Disks}
% allows abbreviation of title, if the full title is too long
% to fit in the running head
%
\author{Wolfgang Brandner\inst{1}
\and Dan Potter\inst{2}
\and Scott S.\ Sheppard\inst{2}
\and Andrea Moneti\inst{3}
\and Hans Zinnecker\inst{4}}
\authorrunning{Wolfgang Brandner et al.\ }
% if there are more than two authors,
% please abbreviate author list for running head
%
%
\institute{European Southern Observatory, Karl-Schwarzschild-Str.\ 2,
D-85748 Garching, Germany; wbrandne@eso.org
\and Institute for Astronomy, University of Hawaii, 2680 Woodlawn Dr., 
Honolulu, HI 96922, USA
\and Institut d'Astrophysique Paris, 98bis Blvd Arago, F-75014 Paris, France
\and Astrophysikalisches Institut Potsdam, An der Sternwarte 16,
D-14482 Potsdam, Germany}

\maketitle              % typesets the title of the contribution

\begin{abstract}
There is strong evidence that the planets in the solar system evolved from a 
disk-shaped solar nebula $\approx$4.56 Gyr ago.  By studying young stars in 
various evolutionary stages, one aims at tracing back the early history of the 
solar system, in particular the timescales for disk dissipation and for the 
formation of planetary systems. We used the VLT \& ISAAC, and ESA's Infrared 
Space Observatory \& ISOCAM to study the circumstellar environment of young 
low-mass stars.

\end{abstract}

\section{Disk-Origin of the Solar System}

The coplanarity of planetary orbits and their moons, and 
the preferentially prograde rotation direction led \cite{kant1755}
and \cite{laplace1796} to the conclusion that the solar system evolved out 
of a flattened disk. Meteoritic evidence from Carbonaceous Chondrites
suggests that the Solar System formed 4.56\,Gyr ago within a time span of
$\le$10\,Myr. Studies of circumstellar disks around young stars aim at
establishing the initial conditions and the exact sequence of events leading 
to the formation of giant and terrestrial planets in the solar system.

\section{How to find Circumstellar Disks? }

Until the late 20th century, the existence of circumstellar disks
around young stars could only be deduced indirectly.
IR and UV excess from young stars were 
interpreted as disk and accretion signatures \cite{lynden1974,adams1987}.
Submm and mm measurements provided the first estimates on disk masses 
and disk lifetimes \cite{beckwith1990,jewitt1994}.
Measurements of the frequency of IR excess, in particular L-band excess
of stars in young clusters reveal that up to 100\% of all young ($\le$1\,Myr) 
stars show evidence for circumstellar disks \cite{haisch2001}.

Physical Parameters of star-disk systems were determined
from their spectral energy distribution (SED) and the
comparison to models \cite{adams1987,bertout1988}. Unfortunately,
solutions derived from this SED fitting turned out to be not
necessarily unique \cite{kenyon1987}. This highlights the need
for spatially resolved observations.
The size of the solar system is $\approx$100\,A.U., which
corresponds to $\le 1''$ at the distance of the nearest star forming regions.
Thus high-spatial resolution observations
are required. We used VLT/ISAAC \cite{moorwood1998}, complemented
by high-sensitivity ISO/ISOCAM \cite{cesarsky1996} observations
for our studies.

\section{High-resolution ANTU/ISAAC survey for edge-on disks}

The main observational difficulty in identifying circumstellar
disks in the visual and near-infrared is that they
are typically 10$^6$ times fainter than the central star.
Circumstellar disks have been detected and spatially resolved
in the Orion HII region as dark silhouettes seen against the bright background
\cite{McCaughrean1996}. Another possibility to detect disks is if
they are oriented close to edge-on (within 5$^\circ$ to 10$^\circ$)
and hence act as natural occulting bars, 
which block out star light \cite{whitney1992,burrows1996,padgett1999}.

\begin{figure}[hbt]
\begin{center}
\includegraphics[width=.82\textwidth]{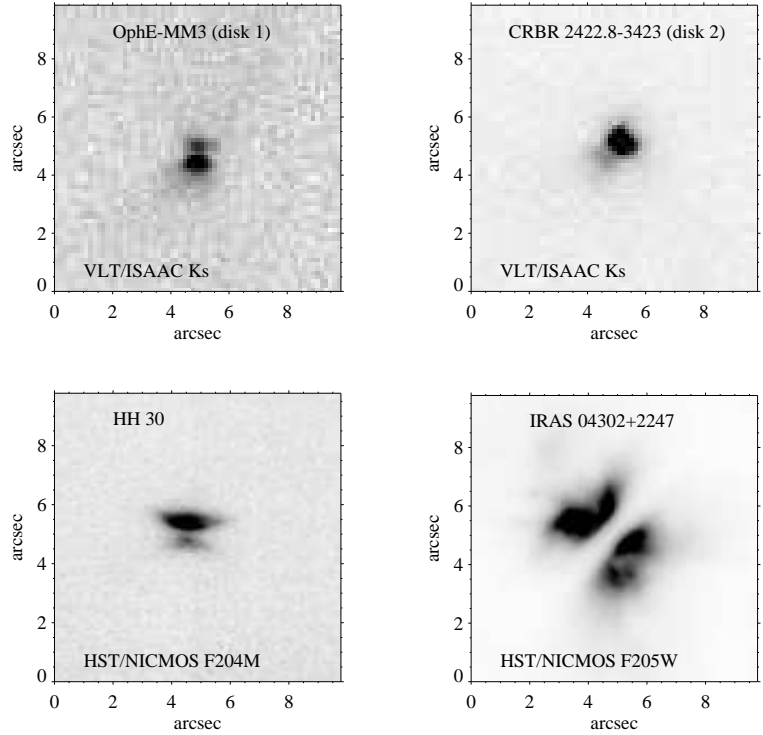}
\end{center}
\caption[]{Top: Edge-on circumstellar disk sources in Ophiuchus detected
with VLT/ISAAC. In comparison to 
isolated disks in Taurus (bottom, shown as observed with HST/NICMOS), the 
Ophiuchus disks and their reflection nebulosities
are more compact. For the disk sources in the Ophiuchus region, North is up 
and East is to the left.}
\label{eps1}
\end{figure}

With the aim to establish a sample of edge-on circumstellar disk sources,
which is suitable for detailed follow-up studies with VLT, VLTI and ALMA,
we employed VLT \& ISAAC to observe southern Class I IRAS sources
associated with faint nebulosities in the optical or NIR.

16 sources were observed with ANTU \& ISAAC in JHKs in April 1999 under
0.35$''$ seeing conditions, which corresponds to a spatial
resolution of 50\,A.U.\ at a distance of 140\,pc, i.e., comparable to
the radius of the Kuiper Belt. The central
dark lane in the Chamaeleon IR Nebula is for the first time nicely
resolved \cite{zinnecker1999}.
Furthermore, two disks seen close
to edge-on could be identified in the $\rho$~Oph star forming region
(Figure \ref{eps1}, \cite{brand2000a}). 
Disk 1 (OphE-MM3) was previously classified as a starless
core or isothermal protostar \cite{motte1998}. It is
located 50$''$ west and north 10$''$ of Elias 2-29 -- one of the most prominent
and most luminous IR sources in Oph. 
Disk 2 (CRBR 2422.8$-$3423), which was originally identified as an
IR source by \cite{comeron1993}, is located 30$''$ west
and 10$''$ south of WLY 2-43.

\begin{figure}[hbt]
\begin{center}
\includegraphics[width=.90\textwidth]{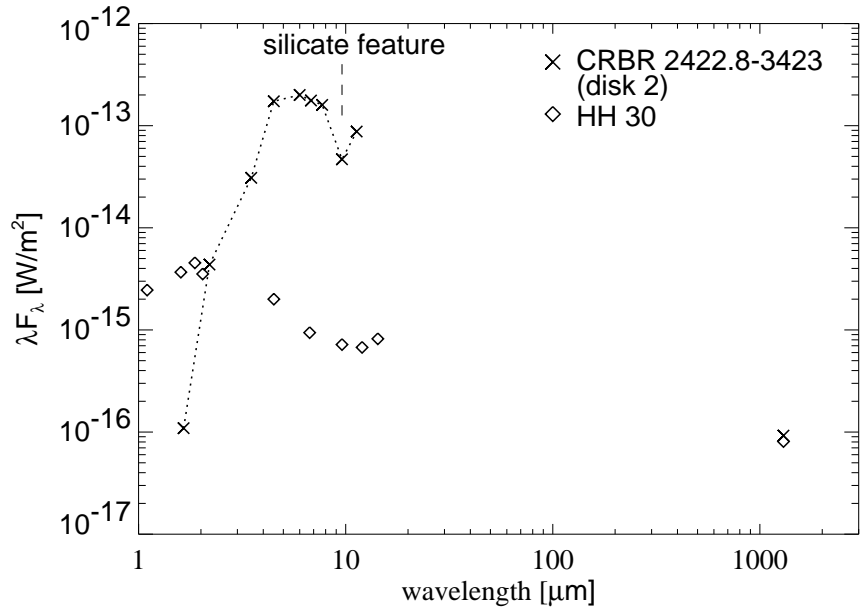}
\end{center}
\caption[]{Spectral energy distribution of CRBR 2422.8$-$3423 (disk 2)
and HH 30. Both sources exhibit about the same flux at 2.2\,$\mu$m and
1.3\,mm. The slightly larger inclination of CRBR 2422.8$-$3423's disk
allows the warm, inner disk to become detectable at NIR to MIR wavelengths.
The dip at 9.6\,$\mu$m can be explained by absorption due to the
silicate feature. The spectral energy distribution of HH 30, whose disk is
seen closer to edge-on than in the case of CRBR 2422.8$-$3423,
is dominated by scattered light out to wavelengths of 10\,$\mu$m.}
\label{eps2}
\end{figure}

The mm measurements by \cite{motte1998} yield disk masses of 
$\approx$0.01\,M$_\odot$, which is also the mass of
the ``minimum solar nebula'' \cite{cameron1962}. 
The disks in the $\rho$\,Oph region
appear to be more compact (100 A.U.\ $\times$ 40 A.U.) than
the majority of edge-on disks in Taurus. Due to the
prevalence of forward scattering, the brightness ratio of the two 
parts of the bipolar reflection nebulosity in each disk directly
yields information on the disk inclination.
Disk 1, which in the NIR is also the fainter of the two disks,
is seen closer to edge-on than disk 2 (inclination angle
85$^\circ$ vs.\ 75$^\circ$).

ISOCAM observations between 3\,$\mu$m and 12\,$\mu$m of disk 2 (inclination
$\approx$75$^\circ$), and of the edge-on disk source HH 30
(inclination $\ge$80$^\circ$) confirm theoretical predictions that
a slight change in the viewing angle of a disk leads to a dramatic difference
in the spectral energy distribution of YSOs 
\cite{boss1996,sonnhalter1995}.
HH 30 and disk 2 have about the same integrated brightness in K
and exhibit the same 1.3\,mm continuum flux. The 
SED of disk 2 is steeply rising between 2.2\,$\mu$m and
6.0\,$\mu$m, and a silicate absorption feature is visible at 9.6\,$\mu$m.
The SED of HH 30 is rather flat, and appears to  be dominated by
scattered light out to a wavelength of 10\,$\mu$m 
\cite{stapelfeldt1999,brand2000a}.

\section{How long do disks survive?}

\begin{figure}[hbt]
\begin{center}
\includegraphics[width=.70\textwidth,angle=270]{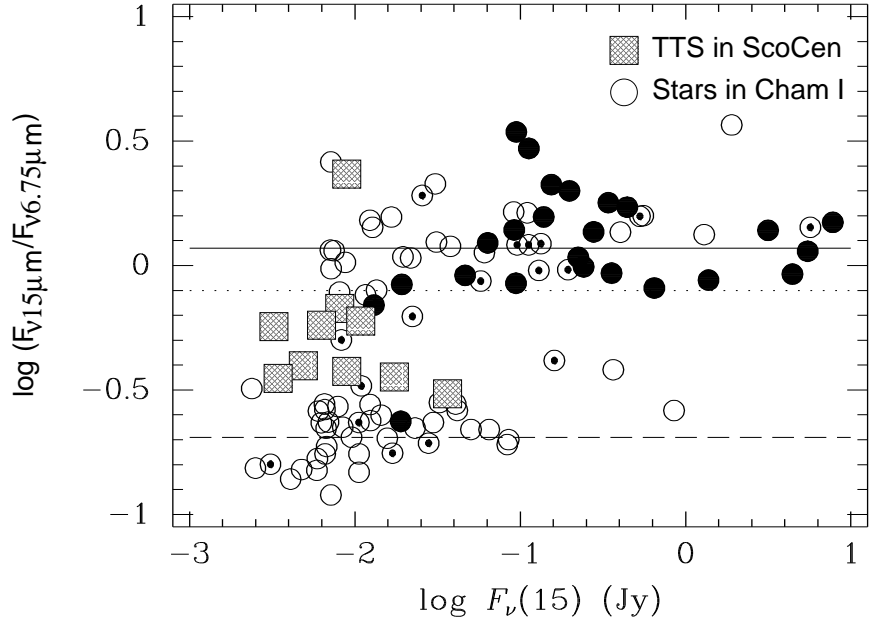}
\end{center}
\caption[]{
Color-magnitude diagram based on ISOCAM observations of
young stars in Chamaeleon (circles) and Scorpius-Centaurus (squares).
Black filled circles indicate
previously known Young Stellar Objects and classical T Tauri stars (CTTS),
circles with a central dot previously known weak-line T Tauri stars (WTTS),
and open circles new sources
detected with ISO by \cite{north1996}. The dashed line indicates
the location of pure stellar photospheres, the dotted and solid
line the location of flat and flared circumstellar disks, respectively,
as predicted by the models from \cite{kenyon1995}.  Post-T\,Tauri 
stars in Scorpius-Centaurus show a spectral index intermediate
between main-sequence stars and CTTS and WTTS in Chamaeleon \cite{brand2000b}.}
\label{eps3}
\end{figure}

Disk survival times and the sequence of events leading to the
formation of planetary systems are closely linked. 
Giant planets in the solar system possess a core of higher density material 
surrounded by a shell of metallic hydrogen and an outer atmosphere.
According to one model, a higher density (rocky) core with a mass of
$\approx$10\,M$_\oplus$ has to form first, before noticeable amounts of nebular
gas can be accreted by the proto-giant planet. Simulations indicate
that at least 10$^6$ yr are necessary to form a 10 M$_\oplus$
rocky core \cite{lissauer1987}, and that another 10$^7$ yr are required
for the 10 M$_\oplus$ core to accrete 300 M$_\oplus$ of nebular
gas.  It is still unknown if massive circumstellar disks
can indeed survive for such an extended period. A second model,
recently reinstated by \cite{boss1998},
suggests that gravitational instability of a protoplanetary disk
leads directly to the formation of a giant gaseous protoplanet on time scales
as short as 10$^3$ yr. The rocky core then forms due to the settling of
dust grains initially acquired, and by further accretion of solid bodies
in the course of the subsequential 0$^5$\,yr.
The difference in timescales for the formation of giant planets in the two 
models provides observational means to decide for or against
either model by studying the  circumstellar environment of stars with ages
$\le$ 15\,Myr.

A sub-mm study of Lindroos binaries with ages between 3 and 150\,Myr
indicates that dust depletion in circumstellar disks occurs within the
first 10\,Myr \cite{jewitt1994}. Similarly, the percentage of sources
with near-infrared and L-band excess in young clusters drops sharply
within the first $\sim$ 6 Myr (e.g., \cite{haisch2001}).

Is this the same period, which defines the ``birth'' of the Solar System
4.56\,Gyr ago?
Do dust and gaseous disk components vanish within 5 to 10\,Myr?
This would imply a potential
timescale problem for the formation of giant planets.

ISO studies of nearby star forming regions with ages of the order of
1\,Myr to 10\,Myr also suggest a gradual decrease of the amount of
circumstellar material (Figure \ref{eps3}), in particular a depletion 
of smaller size dust grains, as stars evolve towards the main-sequence
\cite{north1996,olofsson1999,brand2000b}.

The diminished
infrared excess can be explained by disk dissipation
or by changes in the global dust opacities due
to, e.g., grain growth on timescales of 5 to 15 Myr.

\section{Outlook}

Detailed images of the distribution of scattered light and polarization maps
can now be obtained with adaptive optics systems at 8m class telecopes
(e.g., Hokupa$'$a \& QUIRC at Gemini, and NAOS \& CONICA at VLT). 
The diffraction limit of 50mas in the H-band corresponds to
$<$8 A.U.\ at a distance of 150\,pc, thus the 
inner regions of circumstellar disks (and potential ``protoplanetary
systems'') become resolvable.  In particular
dual imaging methods using a Wollaston prism (in order to 
eliminate speckle noise from unpolarized light) appear to be predestinated
for these kind of studies. Detailed polarization maps combined with
refined theoretical models will enable us to determine physical properties of 
young disks, such as disk geometry, density structure, or dust properties.

High-spectral resolution studies in the infrared (e.g., with CRIRES at VLT),
CRIRES, could probe and (weigh in) the gaseous disks around the post-T Tauri 
stars by searching for H$_2$ features seen in absorption against the stellar 
photosphere. This should provide additional information on the dispersal
time scales for (gaseous) disks and the formation time scales of giant
planets.

%INDEX%%%%%%%%%%%%%%%%%%%%%%%%%%%%%%%%%%%%%%%%%%%%%%%%%%%%%%%%%%%%%%%
% Please check with the editor of your book whether he plans to
% include a "mutual" subject index - if so, please code your entries
% in the standard syntax. For your own purposes you may print your
% "personal" index by using the following commands:
%
%\clearpage
%\addcontentsline{toc}{section}{Index}
%\flushbottom
%\printindex
%%%%%%%%%%%%%%%%%%%%%%%%%%%%%%%%%%%%%%%%%%%%%%%%%%%%%%%%%%%%%%%%%%%%%

\end{document}